\documentclass[a4paper, 11pt]{article}
\usepackage[utf8]{inputenc}
\usepackage[english]{babel}
\usepackage[left=2.5cm, right=2.5cm, top=2cm]{geometry}

\usepackage{hyperref}
\hypersetup{colorlinks=true,linkcolor=red,citecolor=blue,filecolor=black,urlcolor=black,
pdfauthor={Jacopo Mazza}}

\usepackage{graphicx}
\usepackage{caption}
\usepackage{subcaption}

\usepackage{siunitx}
\let\svqty\qty
\usepackage{physics}
\let\qty\svqty
\usepackage{amsmath,amssymb,amsfonts}
\usepackage[scr=boondoxo,scrscaled=1.05]{mathalfa}
\def\be#1\ee{\begin{align}#1\end{align}} 
\def\bse#1\ese{\begin{subequations}#1\end{subequations}}

\usepackage[capitalize]{cleveref}
\Crefname{equation}{Eq.}{Eqs.}
\crefname{equation}{eq.}{eqs.}
\Crefname{figure}{Fig.}{Fig.s} \crefname{figure}{fig.}{fig.s}
\Crefname{table}{table}{tables}
\crefname{subequation}{eqs.}{eqs.}
\labelcrefformat{subequation}{#2(#1)#3}
\crefformat{subequations}{#2(#1)#3}
\crefname{section}{sect.}{sects.}

\usepackage{comment}
\usepackage{soul}
\newcommand{\JM}[1]{\textcolor[RGB]{0,150,50}{\textbf{[JM: #1]}}}

\newcommand{\qu}[1]{`#1'}

\usepackage{authblk}
\newcommand{\IJCLab}{\affil{Université Paris-Saclay, CNRS/IN2P3, IJCLab, 91405 Orsay, France}}

\begin{document}

\title{Black bounces to traversable wormholes from pure gravity in four and higher dimensions}
\author{Jacopo Mazza}
\IJCLab
\date{}

\maketitle
\begin{abstract}
    This article derives static and spherically symmetric wormhole and black bounce spacetimes as vacuum solutions of metric gravitational theories in four and higher dimensions. To this end, I extend previous results on quasi-topological gravities to encompass cases in which the metric depends on two arbitrary functions of one variable --- instead of the usual single function. I then explain how, given (almost) any spherically symmetric metric, one can reverse engineer a well-defined higher-dimensional theory having such a metric as solution. I apply these results to three paradigmatic examples, including the Simpson--Visser black bounce, to showcase the versatility of this method, as well as its limitations. Moreover, I consider the coupling to a Vaidya-like matter source, so as to generate time-varying solutions and investigate what kind of dynamical evolution these theories can account for.   
\end{abstract}

\tableofcontents

    \section{Introduction}

Wormholes are a notoriously popular narrative device in works of science fiction, and a somewhat less popular object of investigation within \qu{scientific non-fiction}.
Their description in formal physical terms dates back, at least, to the celebrated article by Einstein and Rosen \cite{EinsteinParticleProblem1935},\footnote{The birth of the subject is often associated with an article by Flamm \cite{FlammBeitrage1916,FlammRepublicationContributions2015}, which predates \cite{EinsteinParticleProblem1935} by almost twenty years. This association is however contested \cite{GibbonsEditorialNote2015}, and it appears unwarranted.} but it was only after the seminal paper by Morris and Thorne \cite{MorrisWormholesSpacetime1988} that the notion became an integral part of the gravitational physicist's toolbox.
By the present day, wormhole physics has developed into a well-established field, whose state of the art is reviewed e.g.~in \cite{VisserLorentzianWormholes1996,LoboExoticSolutions2007,LoboWormholesWarp2017,BambiAstrophysicalWormholes2021}.
Notably, wormholes are routinely listed among the plausible alternatives to general-relativistic black holes --- almost on par with non-singular black holes, exotic compact objects, and other black hole \qu{mimickers} \cite{Carballo-RubioNonsingularParadigm2025}.

In this respect, a large part of wormholes' appeal is due to the fact that their existence, which is conjectural but at present cannot be completely ruled out, would allow to circumvent some of the issues typically associated with black holes in general relativity, such as the unavoidability of spacetime singularities \cite{Carballo-RubioOpeningPandora2020,Carballo-RubioGeodesicallyComplete2020}.
Arguably, however, so far their analysis has largely been guided by phenomenological considerations, in the sense that wormhole metrics have often \emph{not} been derived from first principles as solutions to self-consistent and \qu{realistic} theories of gravity.
Rather, many such metrics have been proposed and analysed, only to be interpreted \textit{a posteriori} within a given theoretical framework.
This approach is not entirely satisfactory, and it frequently leads to technical issues: in general relativity, wormholes infamously require violations of the energy conditions, though this is probably not as problematic as widely believed \cite{BarceloScalarFields2000,BarceloTwilightEnergy2002}; more to the point, the reverse engineering of a Lagrangian having a specific wormhole metric as solution, be it in general relativity with suitable matter sources or in modified gravity, often requires the fine tuning of parameters \cite{HuangChargedEllis2019,BronnikovFieldSources2022,BronnikovBlackBounces2022,RodriguesSourceBlack2023a,AlencarBlackBounce2024a,AlencarGeneralSpherically2025,PereiraBlackbounceSolution2024,PereiraNewSources2025,SilvaGeneralizedBlackbounces2026}.
Moreover, the possibility of forming a wormhole out of the evolution of regular initial data is speculative \cite{HorowitzCreatingTraversable2019,DaiHowForm2020,ChakrabartiWormholeGeometry2021}; and the stability of such spacetimes is in any case far from guaranteed --- e.g.~\cite{FrancioliniStableWormholes2019}.
Hence, although the possibility of embedding wormholes in (general relativity and) modified gravity theories is well attested and documented --- for some examples, see \cite{HaywardDilatonicWormholes2002,DottiStaticWormhole2007,KorolevKineticGravity2020,BakopoulosTraversableWormholes2022,DeFalcoStaticSpherically2023,RadhakrishnanReviewStable2024,KumarDevelopingFramework2024} and references therein, although this is by no means an exhaustive list ---, their geometrodynamics is still far from being firmly established.

This remark appears particularly enticing in light of a newly emergent approach, whereby the dynamics of four- and higher-dimensional metric theories of gravity is described, in highly symmetric situations, in terms of lower-dimensional theories.
More specifically, within this approach, one exploits e.g.~spherical symmetry to reduce the action of any theory whose \qu{gravitational sector} is described in terms of no field other than the metric to a two-dimensional \qu{effective} scalar--tensor theory --- which, under the assumption that the equations of motion be of order no higher then second in derivatives, must be a Horndeski theory.
To the best of the author's knowledge, this idea can be traced back to \cite{ZiprickQuantumCorrected2010} and to the follow-up works \cite{Louis-MartinezBirkhoffsTheorem1994,TavesModellingEvaporation2014,KunstatterNew2D2016,BarenboimNoDrama2024a,BarenboimEvaporationRegular2025}; although the formalism employed in this article has been laid out most clearly in \cite{Carballo-RubioMasterField2026}.
Notably, it has substantial overlap with that of so-called quasi-topological gravities \cite{BuenoDynamicalFormation2024,BuenoRegularBlack2024a,FrolovRegularBlack2025,FrolovRegularBlack2026,BuenoRegularGeometries2026,KonoplyaQuasinormalModes2026,DubinskyScatteringScalar2026,BuenoRegularBlack2025b,BuenoRegularBlack2026,FrolovVaidyaTypeSolutions2026}, and it has already been subject to a quite remarkable number of extensions and applications \cite{BoyanovRegularVaidya2025,Alonso-BardajiSpacetimeGeometry2024,Alonso-BardajiDynamicalTheory2025,BorissovaEffectiveGeometrodynamics2026,BorissovaRegularBlack2026a,BorissovaAll$2D$2026,BorissovaModifiedFriedmann2026,Borissova$g_ttg_rr1$2026,ArrecheaEffectiveGeometrostatics2026,Carballo-RubioChargingRegular2026,Carballo-RubioBiscalar2026}.

This approach is therefore extremely powerful, and it is natural to wonder whether it could yield new insight on the physics of wormholes in spherical symmetry.
Crucially, although the formalism of \cite{Carballo-RubioMasterField2026} is very general, so far most analyses have focused on a specific subclass of two-dimensional so-called \qu{integrable} theories, which give rise to metrics that can be described, in suitable coordinates, in terms of a single function of one variable.
This choice represents a rather substantial loss of generality, which is justified insofar as it allows to describe a wide variety of physically relevant spacetimes --- including virtually all the most popular non-singular black holes. 
However, it precludes the description of wormholes.\footnote{These remarks refer to vacuum spacetimes. Wormhole solutions in quasi-topological gravities in the presence of suitable matter sources have been discussed in \cite{LiEllisBronnikovWormhole2026}.}

This article aims at (partially) filling this gap.
Specifically, it will extend the formalism of \qu{integrable} Horndeski theories so as to encompass the generic case in which the metric is described by two independent functions.
This extension is thus a substantial contribution to the stream of works \cite{Carballo-RubioMasterField2026,BoyanovRegularVaidya2025,Alonso-BardajiSpacetimeGeometry2024,Alonso-BardajiDynamicalTheory2025,BorissovaEffectiveGeometrodynamics2026,BorissovaRegularBlack2026a,BorissovaAll$2D$2026,BorissovaModifiedFriedmann2026,Borissova$g_ttg_rr1$2026,ArrecheaEffectiveGeometrostatics2026,Carballo-RubioChargingRegular2026,Carballo-RubioBiscalar2026} in its own right.
Moreover, this formalism will be applied to three paradigmatic examples: 
the Simpson--Visser black bounce \cite{SimpsonBlackbounceTraversable2019}, 
a singular deformation of the Schwarzschild metric, 
and the Damour--Solodukhin wormhole \cite{DamourWormholesBlack2007}.
The ensuing discussion will show that some (arguably, most) of the spherically symmetric wormhole metrics that are commonly employed in phenomenological applications can be obtained as solutions of well-defined higher-dimensional modified gravity theories --- modulo some caveats, which are thoroughly described.
Finally, since vacuum spacetimes are necessarily static in this setting, the minimal coupling to a simple time-dependent matter source will be considered --- specifically, a Vaidya-like source ---, so as to showcase what kind of dynamical effects the formalism can naturally account for.

The structure of the article is as follows.
\Cref{sec:2DHorn} reviews known facts on two-dimensional Horndeski theory and introduces the notations used in the rest of the paper.
\Cref{sec:int_reverse} discusses the extension of the notion of \qu{integrable} Horndeski theory necessary to describe generic two-dimensional metrics.
\Cref{sec:Dgeq4} describes the uplift of two-dimensional Horndeski theories to higher dimensions; the description mostly follows \cite{BorissovaEffectiveGeometrodynamics2026,BorissovaRegularBlack2026a,BorissovaAll$2D$2026}, and adds remarks relevant to the extension introduced in the previous section.
\Cref{sec:staicWoH} presents the application of the formalism developed in previous sections to three informative examples of static wormholes.
\Cref{sec:dynamicWoH} extends the discussion introducing a time dependence, dictated by a suitable matter source.
Finally, \cref{sec:concl} reports the author's conclusions.

    \section{\texorpdfstring{$2D$}{2D} Horndeski and its solutions\label{sec:2DHorn}}

Much of the remarkable versatility of the formalism outlined in \cite{Carballo-RubioMasterField2026} can be traced back to the simplicity of Horndeski theory in two dimensions. 
In this section, I review some known facts about this theory, and introduce the notations that I will use for the rest of the article.
Except for possible misprints, my notations should match those of \cite{BoyanovRegularVaidya2025,Carballo-RubioMasterField2026,BorissovaEffectiveGeometrodynamics2026,BorissovaRegularBlack2026a,BorissovaAll$2D$2026,BorissovaModifiedFriedmann2026,Borissova$g_ttg_rr1$2026,ArrecheaEffectiveGeometrostatics2026,Carballo-RubioChargingRegular2026} --- but not entirely those of \cite{Carballo-RubioBiscalar2026}.

    \subsection{Action and equations of motion\label{sec:ActEom}}

As is well known, Horndeski theory \cite{HorndeskiSecondorderScalartensor1974} is the most general theory of a metric $g_{ab}$ and a scalar field $\phi$ with equations of motion of order no higher then second in derivatives \cite{KobayashiHorndeskiTheory2019}. 
In two dimensions, its action can be written as
\be\label{eq:ActHorn}
S_\text{2D Horndeski} \left[ g_{ab}, \phi \right] = \int \dd[2]{x} \sqrt{\abs{g}} \Big\{ &
H_2 \left( \phi, \chi \right) - H_3 \left( \phi, \chi \right) \Box{\phi} + H_4\left( \phi, \chi \right) \mathcal{R}  \nonumber\\
& - 2 H_{4,\, \chi}\left( \phi, \chi \right) \left[ \left(\Box{\phi}\right)^2 - \nabla_a \nabla_b \phi \nabla^a \nabla^b \phi \right] 
\Big\} \, ,
\ee
where the $H_i(\phi, \chi)$ are arbitrary functions of the scalar field $\phi$ and of its kinetic term
\be
\chi := \nabla_a \phi \nabla^a \phi \, ,
\ee
while $\mathcal{R}$ is the (two-dimensional) Ricci scalar.
I am using a notation whereby a variable appearing as subindex indicates differentiation with respect to that variable, so in particular $H_{4,\, \chi} := \partial_\chi H_4$.

The form of the action appearing in \cref{eq:ActHorn} resembles most closely the one familiar from higher dimensions, from which it differs only because it lacks terms specified by an additional function $H_5$, as these terms vanish identically in two dimensions.
That of \cref{eq:ActHorn} is not the most concise writing, however, since the terms specified by $H_4$ can be absorbed into a redefinition of $H_2$ and $H_3$ exploiting integration by parts and dimensionally dependent identities \cite{LovelockDimensionallyDependent1970} --- see e.g.~\cite{Colleaux:2019ckh,TakahashiGeneralized2D2019,ShamsNejatiJackiwTeitelboimGravity2024}.
The action thereby obtained is identical to that of (two-dimensional) kinetic gravity braiding \cite{DeffayetImperfectDark2010}, and it is completely specified in terms of two functions only.
These remarks notwithstanding, the notation of \cref{eq:ActHorn} has become standard, and for this reason it is adopted throughout the entire article.

Irrespective of the specific form in which the action is written, its variation leads to a set of tensorial and scalar equations of motion that read, schematically,
\be\label{eq:eom_vac}
\mathcal{E}_{ab} = 0
\qq{and}
\mathcal{E} = 0 \, .
\ee
The explicit form of these objects is
\be
\label{eq:Eab}
\mathcal{E}_{ab} &= \beta \left[ \nabla_a \nabla_b \phi - g_{ab} \Box{\phi} \right] - \frac{1}{2} \alpha\, g_{ab} + \gamma \nabla_a \phi \nabla_b \phi \, ,\\
\label{eq:Ephi}
\mathcal{E} &= 
\left( \alpha_\phi - 2 \chi \gamma_\phi \right)
+ 2 \left( \beta_\phi - \gamma \right) \Box{\phi}
-2 \gamma_\chi \nabla^a \phi \nabla_a \chi
\nonumber\\
&\phantom{=}
- \beta \mathcal{R} 
+ 2 \beta_\chi \left[ \left( \Box{\phi} 
\right)^2 - \nabla_a \nabla_b \phi \nabla^a \nabla^b \phi \right]\, ,
\ee
where 
\be
\alpha \left( \phi, \chi \right) &:= H_2\left( \phi, \chi \right) + \chi \left[ H_3\left( \phi, \chi \right) - 2 H_{4,\, \phi} \left( \phi, \chi \right) \right]_\phi \, , \label{eq:alpha_def} \\
\beta \left( \phi, \chi \right) &:= \chi \left[ H_3\left( \phi, \chi \right) - 2 H_{4,\, \phi}\left( \phi, \chi \right) \right]_\chi - H_{4,\, \phi} \left( \phi, \chi \right) \label{eq:beta_def} \, , \\
\gamma \left( \phi, \chi \right) &:= \alpha_\chi \left( \phi, \chi \right) - \beta_\phi \left( \phi, \chi \right)\, . \label{eq:gamma_def}
\ee

In practice, $\alpha$ and $\beta$ are arbitrary functions of $\phi$ and $\chi$, inasmuch as the $H_i$ are.
The fact that $\alpha$ and $\beta$ alone suffice to determine the equations of motion completely is a reflection of the redundancy in the writing of \cref{eq:ActHorn}, and it would be obvious in the notation of kinetic gravity braiding. 
Hence, any choice of $\alpha$ and $\beta$ specifies a particular theory within the Horndeski class, and theories corresponding to different $\alpha$ and $\beta$ should be considered as physically inequivalent.
Given a theory specified by $\alpha$ and $\beta$, a particular realisation of its action is given by
\be
H_2 (\phi, \chi) = \alpha(\phi, \chi)
\qq{and}
H_3 (\phi, \chi) = 2 H_{4, \phi}(\phi, \chi) = -2 \beta(\phi, \chi)
\ee
--- cf.~\cite{BorissovaEffectiveGeometrodynamics2026,BorissovaRegularBlack2026a,BorissovaAll$2D$2026};
that is,
\be\label{eq:ActHorn_alphabeta}
S_\text{2D Horndeski} \left[ g_{ab}, \phi \right] = \int \dd[2]{x} \sqrt{\abs{g}} \Big\{ &
\alpha \left( \phi, \chi \right) +2 \beta \left( \phi, \chi \right) \Box{\phi} - \left( \int \dd{\phi} \beta \left( \phi, \chi \right) \right) \mathcal{R}  \nonumber\\
& + 2 \left( \int \dd{\phi} \beta_\chi \left( \phi, \chi \right) \right) \left[ \left(\Box{\phi}\right)^2 - \nabla_a \nabla_b \phi \nabla^a \nabla^b \phi \right] \Big\} \, .
\ee

Further note that the tensorial and scalar equations are not entirely independent, since they are related by the following Bianchi identity:
\be\label{eq:BianchiID}
\nabla^a \mathcal{E}_{ab} + \frac{1}{2} \mathcal{E} \nabla_b \phi = 0 \, .
\ee
Hence, as long as the gradient of the scalar does not vanish, the tensorial equations imply the scalar one.
In this article, unless otherwise specified, $\nabla_a \phi \neq 0$ shall be assumed, and solutions to the tensorial equations alone will therefore be solutions of the full set of equations of motion.

    \subsection{Choice of coordinates and the Birkhoff--Jebsen theorem \label{sec:gauge_Birkhoff}}

The equations of motion of \cref{eq:eom_vac} take a particularly simple form when written in coordinates adapted to the foliation provided by the scalar field $\phi$, i.e.~in which $\phi$ itself plays the role of a coordinate.\footnote{This gauge is reminiscent of the unitary gauge used, for instance, in cosmological --- e.g.~\cite{CheungEffectiveField2008,PiazzaEffectiveField2013} --- and black hole --- e.g.~\cite{FrancioliniEffectiveField2019,MukohyamaEffectiveField2025} --- perturbation theory, as well as in gauge theories.}
This choice is possible as long as $\nabla_a \phi \neq 0$, which is thus another reason why this assumption is made throughout the paper.
Calling $v$ and $r$ the coordinates of the two-dimensional spacetime, this gauge is defined by the conditions
\be\label{eq:phi_gauge}
\pdv{\phi}{v} = 0
\qq{and}
\pdv{\phi}{r} = 1 \, .
\ee
Notably, these conditions do not fix the gauge completely, as they are preserved by coordinate transformations of the form $r \mapsto R = r$ and $v \mapsto V = V(v,r)$.
This residual freedom is sufficient to bring any two-dimensional metric to the following form:
\be\label{eq:metric2D}
g_{ab} \dd{x^a}\dd{x^b} = - f(v,r) \dd{v^2} + 2 h(v,r) \dd{v}\dd{r}\, .
\ee

In this gauge, the tensorial structures entering \cref{eq:Eab} simplify considerably.
In particular,
\be\label{eq:chi_gauge}
\chi (v,r) &= \frac{f(v,r)}{h^2(v,r)} \, .
\ee
Note that, according to this relation, one could trade one of the metric \qu{degrees of freedom} for the kinetic term $\chi(v,r)$.
Hence, the metric could equivalently be parametrised as
\be\label{eq:metric}
g_{ab} \dd{x^a}\dd{x^b} =  h(v,r) \dd{v} \left[ - \chi(v,r) h(v,r) \dd{v} + 2 \dd{r} \right] \, .
\ee
This writing suggests the possibility of reabsorbing the $v$ dependence of the function $h(v,r)$ into a redefinition of $v$, so that the metric depend on this coordinate only through $\chi(v,r)$. 
This would indeed be possible with a coordinate transformation of the form $v=v(V)$, $r=r(V,R)$; however, such a transformation does not preserve the conditions of \cref{eq:phi_gauge}.
That is, the new coordinates would not be adapted to the foliation provided by the scalar field $\phi$.
I shall briefly come back to this point below.

The components of the tensorial equations $\mathcal{E}_{ab} = 0$, instead, read
\be
\mathcal{E}_{rr} &= \alpha_\chi - \beta_\phi - \beta \frac{\partial_r h}{h} = 0  \, , \label{eq:Err} \\
\mathcal{E}_{vr} &= -\frac{1}{2} \alpha h - \frac{1}{2} \beta \left( \frac{ h \partial_r f - 2 f \partial_r h}{h^2} \right) \nonumber\\
&= - \frac{1}{2} \left( \alpha h + \beta h \partial_r \chi \right) = 0 \, , \label{eq:Evr} \\
\mathcal{E}_{vv} + \frac{f}{h} \mathcal{E}_{vr} &= \frac{1}{2} \beta \frac{ h \partial_v f - 2f \partial_v h}{h^2} \nonumber\\
&= \frac{1}{2} \beta h \partial_v \chi = 0 \, . \label{eq:Evv}
\ee
Note that $\alpha$ and $\beta$ appear in these expressions as functions of $v$ and $r$ via \cref{eq:phi_gauge,eq:chi_gauge}.

Given $\alpha$ and $\beta$, one can find solutions to the set of equations $\mathcal{E}_{ab}=0$ following the strategy outlined in \cite{Carballo-RubioMasterField2026}:
\Cref{eq:Evv} entails that the kinetic term $\chi$ cannot depend on  $v$;
\cref{eq:Evr} is then an ordinary differential equation for $\chi(r)$; once this equation has been solved, \cref{eq:Err} can be integrated trivially to give
\be\label{eq:h_traditional}
h(v,r) = h_0(v) \exp(\int \dd{r} \frac{\alpha_\chi - \beta_\phi}{\beta})\, .
\ee
Here, $h_0(v)$ is an arbitrary function of $v$ that appears as an integration constant.
However, this function can always be reabsorbed via a redefinition of $v$, hence it can be set to one without loss of generality.

Note that, given $\alpha$ and $\beta$, the solution to the equations of motion is unique.
Moreover, the resulting metric is automatically independent from $v$, and the vector $\xi^a\partial_a = \partial_v$ is therefore a Killing vector.
If such a Killing vector is timelike, at least in some domain of $r$, the metric can rightfully be called static.
Hence, all these theories satisfy a Birkhoff--Jebsen theorem \cite{JebsenGeneralSpherically2005,BirkhoffRelativityModern1923,DeserIntroductionJebsens2005,VojeJohansenDiscoveryBirkhoffs2006}.
Incidentally, this entails that in order to allow for the metric to depend on $v$, two-dimensional Horndeski theory must be deformed, e.g.~by coupling it to matter.

    \section{\qu{Integrable} theories and their reverse engineering\label{sec:int_reverse}}

To date, most part of the literature has focused on the particular case $h(v,r) = 1$.\footnote{If the metric were written in Schwarzschild gauge, the condition equivalent to $h=1$ would be $g_{tt}g_{rr} = -1$ --- cf.~\cite{Borissova$g_ttg_rr1$2026}.}
According to \cref{eq:Err}, $h = 1$ is possible if the functions $\alpha$ and $\beta$ satisfy the \qu{integrability condition}
\be\label{eq:res_int_cond}
\alpha_\chi - \beta_\phi = 0 \, ,
\ee
which is guaranteed to hold if $\alpha$ and $\beta$ can be derived from a \qu{potential} $\Omega(\phi, \chi)$ according to
\be
\alpha = \Omega_\phi
\qq{and}
\beta = \Omega_\chi \, .
\ee
For this reason, much of the literature now identifies the case $h=1$ with the requirement that there exists a potential $\Omega$, and refers to the subclass of two-dimensional Horndeski theories for which the condition of \cref{eq:res_int_cond} is identically satisfied as \qu{integrable theories}.

This particular case is surprisingly rich and it allows, among other things, to obtain all the most popular non-singular black holes --- e.g.~Bardeen \cite{BardeenNonsingularGeneral1968}, Frolov \cite{FrolovSphericallySymmetric1981}, Dymnikova \cite{DymnikovaVacuumNonsingular1992}, Hayward \cite{HaywardFormationEvaporation2006} --- as solutions of well-defined two-dimensional Horndeski theories. 
In fact, virtually any metric that can be described in terms of a single function $f(r)$ may be obtained in this way.

However, limiting the analysis to this subclass of theories represents a substantial restriction: for instance, as I will describe in great detail below, it makes it impossible to account for wormholes.
Moving past this restriction thus seems necessary.
Somewhat surprisingly, the required generalisation entails no substantial increase in complexity: as I will show momentarily, the formalism of integrable theories can be extended quite trivially to the case $h \neq 1$. 
In fact, the two cases appear so similar that distinguishing between them might be superfluous, and the terminology \qu{integrable theories} might need to be revised.

    \subsection{A theory for every solution \label{sec:theorysol}}

To treat the case in which the function $h(r)$ in the metric of \cref{eq:metric2D} is generic, introduce an arbitrary function $\eta(\phi)$, assumed not to be identically zero, and consider the quantity
\be
\omega := \alpha(\phi, \chi) \eta(\phi) \dd{\phi} + \beta(\phi, \chi) \eta(\phi) \dd{\chi} \, ,
\ee
which is a one-form in the field space $\{ \phi, \chi \}$. 
Requiring that $\omega$ be closed, i.e.~that
\be\label{eq:omega_closed}
\dd{\omega} = - \left[ \alpha_\chi \eta - \left( \beta \eta \right)_\phi \right] \dd{\phi} \wedge \dd{\chi} = 0 \, ,
\ee
is equivalent to writing \cref{eq:Err} as
\be
\mathcal{E}_{rr} = \beta h \left[ \frac{\eta_\phi}{\eta} - \frac{\partial_r h}{h} \right] = 0 \, .
\ee
Assuming $\beta \neq 0$, and recalling that in this gauge $\phi$ is identified with the coordinate $r$, the unique solution to this equation is
\be
h(v,r) 
&= h_0(v) \eta(r)\, ,
\ee
where the integration function $h_0(v)$ can be set to one without loss of generality, as before --- cf.~\cref{eq:h_traditional}.
Hence, the field-space function $\eta(\phi)$ determines the shape of the solution's $h(r)$, and vice versa.

Moreover, requiring that $\dd{\omega} = 0$ in an open subdomain of $\{ \phi, \chi \}$ entails, by Poincaré's lemma, that $\omega = \dd{\Omega}$ for some $\Omega(\phi, \chi)$.
In such a case, one necessarily has
\be\label{eq:Omega_def}
\alpha(\phi, \chi) = \frac{\Omega_\phi (\phi, \chi) }{\eta(\phi)}
\qq{and}
\beta(\phi, \chi) = \frac{\Omega_\chi (\phi, \chi) }{\eta(\phi)} \, .
\ee
Accordingly, \cref{eq:Evr} becomes
\be\label{eq:Evr_Omega}
\mathcal{E}_{vr} &= - \frac{1}{2} \left[ \Omega_\phi + \Omega_\chi \partial_r \chi \right] \nonumber\\
&= - \frac{1}{2} \dv{}{r} \Omega(r,\chi(r)) = 0 \, ,
\ee
which determines the solution $\chi(r)$ by the requirement that $\Omega$ be constant when evaluated on shell.
Note that this is an algebraic equation for $\chi(r)$.
Moreover, the solution will depend continuously on the (arbitrary) integration constant, i.e.~on the primary hair, that represents the value of $\Omega$ on shell --- call such a constant $4M$.


To summarise: for any choice of functions $\Omega(\phi, \chi)$ and $\eta(\phi)$, there exists a two-dimensional Horndeski theory, specified by $\alpha$ and $\beta$ as per \cref{eq:Omega_def}, whose unique solution has $h(r)$ and $\chi(r)$ given by
\be
h(r) = \eta(r) 
\qq{and}
\Omega(r, \chi(r)) = 4M \, .
\ee

Correspondingly, given a metric specified by a function $f(r; M)$, which depends on a parameter $M$, and a function $h(r)$, which does \emph{not} depend on $M$, one can reverse engineer a two-dimensional Horndeski theory of which the metric is the unique solution.
To achieve this goal, one may simply invert the relation between $f$ and $M$ to write $M(r, f)$; the theory is then specified by $\Omega(\phi, \chi) = 4 M (\phi, h^2(\phi) \chi)$ and $\eta(\phi) = h(\phi)$.
Such a reverse engineering is therefore unique, as long as $f$ is a bijection when regarded as a function of $M$.

I shall discuss some explicit examples of solutions and of their respective theories in \cref{sec:staicWoH}.
Here, I only wish to stress that the construction presented here is extremely general, as a large swath of two-dimensional metrics of physical interest can be generated in this way. 
In fact, the only restriction on the kind of metrics compatible with this construction is that they should depend on only one integration constant --- more specifically, $f(r)$ must depend on only one integration constant, while $h(r)$ must not depend on said constant. 
Any other parameter that might appear in the metric must be present already at the level of the action, and must therefore be interpreted as a coupling constant.

Note also that this construction is identical to the one of the so-called \qu{integrable case}, which is recovered if $\eta(\phi)$ is set to one. 
For this reason, I suggested above that distinguishing between the case $h=1$ and $h\neq1$ might be superfluous.
Yet, a suitably enlarged notion of \qu{integrable theories} might still be meaningful, due to a subtlety over which I glossed above and that I will further discuss below.

    \subsection{\qu{Extended integrability} of \texorpdfstring{$2D$}{2D} theories \label{sec:int2D}}

In the construction presented in \cref{sec:theorysol}, the existence of the \qu{potential} $\Omega$ and the writing of \cref{eq:Omega_def} require that the one-form $\omega$ be closed everywhere on the space $\{ \phi, \chi \}$ --- or, at least, in an open set therein.
A solution to the equations of motion, on the other hand, is a curve in the space $\{\phi, \chi\}$, which is not an open subset in the sense needed for Poincaré's lemma to hold.\footnote{Technically, what constitutes an open set depends on the choice of topology. What I mean is that any neighbourhood around any point on the curve contains points which do not belong to the curve, and at which $\dd{\omega}$ is not required to be zero. Hence one cannot conclude that $\omega = \dd{\Omega}$. }
Hence, the existence of $\Omega$ and the writing of \cref{eq:Omega_def} are not, technically speaking, a consequence of the equations of motion.
Rather, what the equation of motion $\mathcal{E}_{rr} = 0$ implies is merely that $\dd{\omega} = 0$ \emph{on shell}.

Hence, the requirement that $\Omega$ exists and the writing of \cref{eq:Omega_def} holds implements an \emph{off shell} constraint that specifies a proper subset of two-dimensional Horndeski theories.
This constraint is encoded in \cref{eq:omega_closed}, which should be read as the requirement that there exists $\eta(\phi)$ such that the equation holds.
\Cref{eq:omega_closed} can be integrated straightforwardly to give
\be\label{eq:eta_fromalphabeta}
\eta(\phi) = \exp( \int \dd{\phi} \frac{\alpha_\chi - \beta_\phi}{ \beta} ) \, ,
\ee
up to an integration constant [cf.~\cref{eq:h_traditional}]; since the left-hand side is assumed not to depend on $\chi$, $\alpha$ and $\beta$ must be such that
\be\label{eq:ext_int_cond}
\left( \frac{\alpha_\chi - \beta_\phi}{ \beta} \right)_\chi = 0\, ,
\ee
which must hold, in principle, for every $\phi$ and every $\chi$.

It seems reasonable to refer to theories in this subset as \qu{integrable theories}, and to the constraint of \cref{eq:ext_int_cond} as \qu{integrability condition}.
Note, however, that these arguments hold irrespective of whether $\eta(\phi) = 1$ or not.
Hence, the notion of \qu{integrability} introduced here extends the one adopted in e.g.~\cite{BorissovaEffectiveGeometrodynamics2026,BorissovaRegularBlack2026a,BorissovaAll$2D$2026} and references therein.
In particular, the property of a theory being integrable is distinct from that of having a solution such that $h(r) = 1$. 

A few comments are in order, at this point, concerning the interpretation of the function $\eta(\phi)$.
The writing of \cref{eq:Omega_def} might suggest that $\eta(\phi)$ should be interpreted as a \qu{new potential}, on par with $\Omega$.
This point of view is corroborated by the considerations on the reverse engineering of a theory starting from its solution, since $\eta$ is in one-to-one correspondence with the metric function $h(r)$.
However, this interpretation is not entirely correct, as \cref{eq:eta_fromalphabeta} attests.
Indeed, $\alpha$ and $\beta$ can still be expressed in terms of $\Omega$ only, which rightfully deserves to be called \qu{potential} --- exactly as in the particular case described in previous literature.
What distinguishes the generic case discussed here from that particular case is the functional form of the relation between $\alpha$ and $\beta$, on one side, and (the derivatives of) $\Omega$, on the other: while in the particular case, such relation is linear, in the generic case it involves non-linearities --- so much so that, in practice, it might not be possible to write $\alpha$ and $\beta$ \emph{explicitly} in terms of $\Omega$ alone.

To appreciate this subtlety in finer detail, consider the possibility of writing the action of an integrable theory in terms of $\Omega$ only.
Doing this would require replacing $\alpha$ and $\beta$ in, say, \cref{eq:ActHorn_alphabeta} with their expressions in terms of (the derivatives of) $\Omega$.
According to \cref{eq:Omega_def}, this leads to
\be\label{eq:ActOmegaeta}
S_\text{integrable} \left[ g_{ab}, \phi \right] &= \int \dd[2]{x} \sqrt{\abs{g}} \left\{
\frac{\Omega_\phi \left( \phi, \chi \right) }{\eta(\phi)} + 2 \frac{ \Omega_\chi \left( \phi, \chi \right) }{\eta (\phi) } \Box{\phi} - \left( \int \dd{\phi} \frac{ \Omega_\chi \left( \phi, \chi \right) }{\eta(\phi)} \right) \mathcal{R}  \right. \nonumber\\
& \left. + 2 \left( \int \dd{\phi} \frac{ \Omega_{\chi \chi} \left( \phi, \chi \right) }{\eta(\phi)} \right) \left[ \left(\Box{\phi}\right)^2 - \nabla_a \nabla_b \phi \nabla^a \nabla^b \phi \right] 
\right\} \, .
\ee
If such a theory arises via reverse engineering starting from a given metric, the expression above is perfectly viable, since the function $\eta$ is unambiguously known.
On the other hand, technically, the instance of $\eta$ in \cref{eq:ActOmegaeta} is only a formal writing, as one should replace it with its explicit expression in terms of $\alpha$ and $\beta$, hence of $\Omega$.
The resulting action is still well defined, as in the particular case treated in previous literature, but it is considerably less wieldy and, likely, it can only be written in an implicit form.

The upshot of this discussion is that the introduction of the \qu{free} function $\eta(\phi)$ is instrumental in the reverse engineering of a theory starting from its solution.
However, in a more abstract sense, $\eta$ is merely a notation, and its introduction is arguably overkill.
Indeed, integrable theories, in the extended sense introduced here, can be taken to be defined by the (extended) integrability condition of \cref{eq:ext_int_cond}; these theories are such that $\alpha$ and $\beta$ can be derived from a potential $\Omega$, though in a rather involved way; at any rate, their solutions can be determined following the general strategy outlined in \cite{Carballo-RubioMasterField2026}.

\bigskip

I close this section with some simple yet suggestive remarks concerning the mutual relation between theories belonging to this extended integrable class.
These remarks will further showcase the versatility of this formalism, and provide useful results to which I will come back in \cref{sec:staicWoH}.
Specifically, I will assume that a metric written as in \cref{eq:metric2D}, i.e.~specified by two functions $f(r)$ and $h(r)$, is the solution to an integrable theory with associated potential $\Omega$; and I will consider two simple deformations of this scenario.

As a first deformation, consider a theory specified by the same potential $\Omega$ but corresponding to an $\tilde{\eta}$ different from $\eta$.
The solution of such a deformed theory will have the same profile of $\chi(r)$ as the original solution.
However, the metric will amount to a deformation of the metric in \cref{eq:metric2D} with $\tilde{h}(r)=\tilde{\eta}(r)$, instead of $h(r)$, and
\be
\tilde{f}(r) = f(r) \frac{\tilde{\eta}^2(r)}{\eta^2(r)} \, ,
\ee
instead of $f(r)$.
The functions $\tilde{\alpha}$ and $\tilde{\beta}$ that specify the new theory, instead, will simply be rescaled versions of the original $\alpha$ and $\beta$.
It would certainly be interesting to further investigate the effect of this deformation at the level of the Lagrangian, but this would exceed the scope of this article and I therefore leave it for future work.

As a second deformation, consider a metric whereby $h(r)$ is replaced with some $\tilde{h}(r)$, while $f(r)$ is left untouched.
One can reverse engineer an integrable theory of which the deformed metric is the solution, thereby obtaining a \qu{deformed} theory specified by some $\tilde{\Omega}$. 
This $\tilde{\Omega}$ can be expressed entirely in terms of $\Omega$ in a rather simple way.
Indeed, if the original theory has $\Omega\left( \phi, \chi \right) = 4M \left( \phi, \chi h^2(\phi) \right)$, its deformation will have
\be
\tilde{\Omega} \left( \phi, \chi \right) &= 4M \left( \phi, \chi \tilde{h}^2 (\phi) \right) \nonumber\\
&= \Omega \left( \phi, \chi \tilde{h}^2 (\phi) / h^2(\phi) \right)\, .
\ee
Therefore
\be
\tilde{\Omega}_\chi \left( \phi, \chi \right) &= \frac{\tilde{h}^2(\phi)}{h^2(\phi)} \Omega_\chi \left( \phi, \chi \tilde{h}^2/h^2 \right) \, , \\
\tilde{\Omega}_\phi \left( \phi, \chi \right) 
&= \Omega_\phi \left( \phi, \chi \tilde{h}^2/h^2 \right) + \chi \Omega_\chi \left( \phi, \chi \tilde{h}^2/h^2 \right) \dv{}{\phi} \left[ \frac{\tilde{h}^2(\phi)}{h^2(\phi)} \right] \nonumber\\
&= \Omega_\phi \left( \phi, \chi \tilde{h}^2/h^2 \right) + \chi \tilde{\Omega}_\chi \left( \phi, \chi \right) \dv{}{\phi} \log( \frac{\tilde{h}^2(\phi)}{h^2(\phi)} )
\, .
\ee
Correspondingly, according to \cref{eq:Omega_def},
\be
\tilde{\alpha} (\phi, \chi) &= \frac{h(\phi)}{\tilde{h}(\phi)} \left\{ \alpha \left( \phi, \chi \tilde{h}^2/h^2 \right) + \chi \beta \left( \phi, \chi \tilde{h}^2/h^2 \right) \dv{}{\phi} \left[ \frac{ \tilde{h}^2 (\phi) }{ h^2 (\phi) } \right] \right\}\, , \label{eq:alpha_deform} \\
\tilde{\beta} (\phi, \chi) &= \frac{\tilde{h}(\phi)}{h (\phi) } \beta \left( \phi, \chi \tilde{h}^2/h^2  \right) \, . \label{eq:beta_deform}
\ee
As before, it would be interesting to investigate the effect of this deformation at the level of the action, but this is beyond the scope of this work.

    \section{Uplifting to \texorpdfstring{$D \geq 4$}{D ≥ 4} \label{sec:Dgeq4}}

In the spirit of \cite{BoyanovRegularVaidya2025,Carballo-RubioMasterField2026,BorissovaEffectiveGeometrodynamics2026,BorissovaRegularBlack2026a,BorissovaAll$2D$2026,BorissovaModifiedFriedmann2026,Borissova$g_ttg_rr1$2026,ArrecheaEffectiveGeometrostatics2026,Carballo-RubioChargingRegular2026,Carballo-RubioBiscalar2026}, the two-dimensional Horndeski theory of \cref{eq:ActHorn} is meant to provide an effective description for the dynamics of modified gravity theories of the form
\be\label{eq:ActDgeq4}
S\left[ g_{\mu\nu} \right] = \int \dd[D]{x} \sqrt{ \abs{ \det(g_{\mu \nu})} } \mathcal{L} \left( g^{\mu \nu}, R_{\mu \nu \rho \sigma}, \nabla_\mu \right) \, ,
\ee
in $D \geq 4$ spacetime dimensions, constructed out of: the $D$-dimensional metric $g_{\mu\nu}$, the covariant derivative $\nabla_\mu$ compatible with $g_{\mu \nu}$, the associated Riemann tensor $R_{\mu \nu \rho \sigma}$; and nothing else.

Such an effective description is supposed to be valid for spacetimes that have the structure of a $2+(D-2)$ warped product, i.e.~spacetimes whose metric may be written as 
\be\label{eq:metricDgeq4}
g_{\mu \nu} (x^\alpha) \dd{x^\mu} \dd{x^\nu} = g_{ab}(x^c) \dd{x^a} \dd{x^b} + \phi^2(x^c) \gamma_{ij} \dd{\theta^i} \dd{\theta^j} \, .
\ee
Here, $\gamma_{ij}$ is the metric of the $(D-2)$-dimensional sections, which for definiteness I will assume to be spheres; $g_{ab}$ is the metric of the two-dimensional sections; and $\phi$ is a warping function, which is a scalar on the two-dimensional sections.

When evaluated on spacetimes of the form of \cref{eq:metricDgeq4}, the action of \cref{eq:ActDgeq4} gives rise to an effective two-dimensional action for the two-dimensional metric $g_{ab}$ and the two-dimensional scalar $\phi$. 
If the $(D-2)$-dimensional sections are symmetric under the action of a group that is compact, the principle of symmetric criticality \cite{PalaisPrincipleSymmetric1979,FelsPrincipleSymmetric2002,DeserShortcutsHigh2003} ensures that the equations of motion derived from the effective action are equivalent to the restriction of the equations of motion derived from the original $D$-dimensional action.

If one assumes that the symmetry-reduced equations of motion are of order no higher then second in derivatives, the effective action must be within the Horndeski class of \cref{eq:ActHorn}.
In such a case, one can write
\be\label{eq:Act_Warp}
\eval{ S \left[ g_{\mu\nu} \right] }_\text{warped product} 
&= \mathcal{A}_{D-2} \, S_\text{2D Horndeski} \left[ g_{ab}, \phi \right] \, ,
\ee
where $\mathcal{A}_{D-2}$ is the (constant) volume of the $(D-2)$-dimensional sections, as defined by $\gamma_{ij}$, while $S_\text{2D Horndeski}$ is as in \cref{eq:ActHorn}.
This reduction is the gist of the effective approach proposed in \cite{Carballo-RubioMasterField2026}.

Incidentally, note that the gauge choice of \cref{sec:gauge_Birkhoff}, implemented by the conditions of \cref{eq:phi_gauge}, entails using the warping function $\phi$ as a coordinate for the two-dimensional sections.
The warping function measures the surface area of the $(D-2)$-dimensional sections as immersed submanifolds of the $D$-dimensional spacetime: in $D=4$ and in spherical symmetry, $\phi$ is called \qu{areal radius}, which justifies the use of the letter $r$ for the corresponding coordinate.
As mentioned above, this choice of coordinates is technically possible only provided $\partial_r \phi \neq 0$.

The fact that \emph{generic} two-dimensional Horndeski theories can be obtained as symmetric reductions of this type is not trivial and has been shown in \cite{BorissovaAll$2D$2026}, building upon \cite{BorissovaEffectiveGeometrodynamics2026,BorissovaRegularBlack2026a}.
Generally, the action of such higher-dimensional theories will be a nonpolynomial function of the scalars that can be constructed out of the metric, the Riemann tensor, and covariant derivatives thereof. 

More specifically, given a $2D$ Horndeski theory expressed as in \cref{eq:ActHorn}, a $D$-dimensional theory whose symmetric reduction yields said theory can be written as 
\be\label{eq:ActDgeq4_from2D}
S_\text{uplift} \left[ g_{\mu \nu} \right] & = \int \dd[D]{x} \sqrt{\abs{\det(g_{\mu \nu})} } 
\left( \mathcal{I}_\phi \right)^{D-2} \Big\{
H_2 \left( \mathcal{I}_\phi, \mathcal{I}_\chi\right) - H_3\left( \mathcal{I}_\phi, \mathcal{I}_\chi\right) \mathcal{I}_{\Box{\phi}} + H_4\left( \mathcal{I}_\phi, \mathcal{I}_\chi\right) \mathcal{I}_\mathcal{R}  \nonumber\\
& -2  H_{4,\, \mathcal{I}_\chi }\left( \mathcal{I}_\phi, \mathcal{I}_\chi\right) \left[ \left(\mathcal{I}_{\Box{\phi}}\right)^2 - \mathcal{I}_{\nabla_a \nabla_b \phi \nabla^a \nabla^b \phi} \right] 
\Big\} \, ,
\ee
where the $\mathcal{I}_I$ are $D$-dimensional scalars that coincide with their subscript when evaluated on warped-product spacetimes of the form of \cref{eq:metricDgeq4}.
For instance, $\mathcal{I}_\phi$ is a quantity such that $\eval{\mathcal{I}_\phi}_{\text{warped product}} = \phi$.
Explicit realisations of these $\mathcal{I}_I$ in $D=4$ can be found in \cite{BorissovaEffectiveGeometrodynamics2026,BorissovaRegularBlack2026a,BorissovaAll$2D$2026}.

It is worth highlighting that \cref{eq:ActDgeq4_from2D} provides a way of uplifting any two-dimensional Horndeski theory to a higher-dimensional \qu{purely metric} theory.
Such an uplift can hardly be expected to be unique, as testified by the ambiguity with which the scalars $\mathcal{I}_I$ are defined.
Indeed, these quantities are only required to yield the desired expressions when evaluated on warped-product spacetimes, and they could therefore admit multiple realisations.

Generically, the action of \cref{eq:ActDgeq4_from2D} involves derivatives of the Riemann tensor, in the sense that at least some of the $\mathcal{I}_I$ will be constructed in terms of said derivatives.
The requirement that such derivatives be absent is a constraint on the form of the higher-dimensional action of \cref{eq:ActDgeq4_from2D}, which in turn identifies a proper subset of two-dimensional Horndeski theories via dimensional reduction.
Said constraint is expressed in terms of the functions $H_i$ as the requirement that they can be written as
\be\label{eq:H_NoDer}
H_2(\phi, \chi) = \phi^{D-2} \mathbb{H}_2(\psi) \, ,
\quad
H_3(\phi, \chi) = \phi^{D-3} \mathbb{H}_3(\psi) \, ,
\quad
H_4(\phi, \chi) = \phi^{D-2} \mathbb{H}_4(\psi) \, ,
\ee
where the $\mathbb{H}_i$ are arbitrary functions of the only variable
\be
\psi := \frac{1-\chi}{\phi^2} \, .
\ee
(I am taking the $(D-2)$ dimensional sections to be spheres: in general, the definition of $\psi$ depends on the curvature of such sections.)
Roughly speaking, the reason for this highly constrained form is that (nonderivative) curvature invariants constructed out of the $D$-dimensional Riemann tensor depend on $\chi$ only through the combination $\psi$.
Correspondingly, the equations of motion of such theories will be determined by $\alpha$ and $\beta$ having the specific form
\be\label{eq:alpha_NoDer}
\alpha (\phi, \chi) &= \phi^{D-2} \mathbb{G}_1(\psi) + \phi^{D-4} \mathbb{G}_2(\psi) \, ,\\
\label{eq:beta_NoDer}
\beta (\phi, \chi) &= \phi^{D-3} \mathbb{G}_3(\psi) + \phi^{D-5} \mathbb{G}_5(\psi) \, , 
\ee
where the $\mathbb{G}_i$ are particular combinations of the $\mathbb{H}_i$, whose explicit expression can be found in \cite[eqs.~(97)--(100)]{BorissovaAll$2D$2026}.
Notably, these functions are arbitrary inasmuch as the $\mathbb{H}_i$ are.
Hence, determining whether a given two-dimensional theory can be uplifted to a higher-dimensional theory that does not involve derivatives of the Riemann is fairly straightforward: roughly speaking, if its equations of motion can be written in terms of the combination $\psi$, then such an uplift is possible.

Given the relevance of integrable two-dimensional Horndeski theories for physical applications, a closer look at their uplift to higher dimensions seems warranted.
The authors of \cite{BorissovaEffectiveGeometrodynamics2026,BorissovaRegularBlack2026a,BorissovaAll$2D$2026} addressed this issue for the restrictive notion of integrability related to $h=1$.
They found that $2D$ integrable theories can be obtained from the symmetric reduction of $D\geq4$ theories with and without derivatives; concerning theories without derivatives, in particular, they found that such theories can be both polynomial and not polynomial in $D \geq 5$, but they must be not polynomial in $D=4$.
As a result of this analysis, \cite{BorissovaAll$2D$2026} proposes to use the term \qu{quasi-topological gravities} to refer collectively to all the $D\geq4$ theories whose symmetric reduction is an \qu{integrable} Horndeski theory, in the restrictive sense.

The discussion in the previous section suggests that this nomenclature might be unnecessarily restrictive, as the notion of \qu{quasi-topological gravities} would deserve to be extended along with that of \qu{integrability}.
Namely, it seems reasonable to call \qu{quasi-topological gravities} all the higher-dimensional theories whose symmetric reduction yields two-dimensional theories that are integrable in the extended sense of \cref{eq:ext_int_cond}.
Such an extension falls well within the domain of validity of the formalism developed in \cite{BorissovaEffectiveGeometrodynamics2026,BorissovaRegularBlack2026a,BorissovaAll$2D$2026}, and it can be carried out with only a relatively mild increase in complexity.

As already discussed in the previous section, such an increase in complexity is ultimately related to the non-linearity of the extended integrability condition --- \cref{eq:ext_int_cond} instead of \cref{eq:res_int_cond} ---, and to the corresponding non-linear dependence of $\alpha$ and $\beta$ on (the derivatives of) $\Omega$.
However, from the point of view of the reverse engineering of a theory starting from a known metric, these complications appear rather inconsequential.

\bigskip

In closing this section, I point out that the effective approach sketched herein obviously applies to general relativity in four and higher spacetime dimensions.
The $D$-dimensional Ricci scalar can be decomposed as \cite{BorissovaAll$2D$2026}
\be\label{eq:RicciDgeq4}
\eval{R^{(D)}}_\text{warped product} = \mathcal{R} - 2 (D-2) \frac{\Box{\phi}}{\phi} + (D-3)(D-2) \frac{1-\chi}{\phi^2} \, .
\ee
Hence, focusing on $D=4$ for definiteness, the Einstein--Hilbert action gives rise to the following two dimensional effective action
\be\label{eq:ActGR}
S_\text{GR} \left[ g_{ab}, \phi \right] = \int \dd[2]{x} \sqrt{\abs{g}} \phi^2 \left[
\frac{2(1-\chi)}{\phi^2} - 4 \frac{\Box{\phi}}{\phi} + \mathcal{R}
\right] 
\, ,
\ee
which is a two-dimensional Horndeski action of the form of \cref{eq:ActHorn_alphabeta} with \cite{Carballo-RubioMasterField2026}
\be\label{eq:alphabetaGR}
\alpha \left( \phi, \chi \right) = \alpha_\text{GR} \left( \phi, \chi \right) := 2(1-\chi)
\qq{and}
\beta\left( \phi, \chi \right) = \beta_\text{GR} \left( \phi, \chi \right) := -2\phi \, .
\ee
Note that $\left( \alpha_\text{GR} \right)_\chi - \left( \beta_\text{GR} \right)_\phi = 0$, hence general relativity gives rise to a two-dimensional Horndeski that is integrable in the restrictive sense.
The respective potential is
\be\label{eq:OmegaGR}
\Omega\left( \phi, \chi \right) = \Omega_\text{GR} \left( \phi, \chi \right) := 2 \phi (1-\chi) \, ;
\ee
evaluating it on shell, i.e.~on the Schwarzschild solution, gives
\be
\Omega_\text{GR} \left( r, 1- \frac{2M}{r} \right) = 4M\, ,
\ee
as it must.
Note that $\alpha_\text{GR}$ and $\beta_\text{GR}$ can be written as in \cref{eq:alpha_NoDer,eq:beta_NoDer}, and correspondingly $S_\text{GR}$ is of the form of \cref{eq:H_NoDer}, as they must, since the Einstein--Hilbert action obviously does not involve derivatives of the Riemann tensor.
Interestingly, however, it does not at all appear obvious that the uplift of \cref{eq:ActGR} yields \emph{only} the Einstein--Hilbert action.

    \section{Static wormholes and black bounces \label{sec:staicWoH}}

Though the previous section has steered the discussion towards higher-dimensional spacetimes, which are arguably more relevant from a physical point of view, up to this point the treatment has remained fairly abstract.
In this and the next section, instead, I will specify the general results presented above to some notable examples.
In particular, these examples will consists in higher-dimensional metrics that describe wormholes and so-called black bounces.

Specifically, I will consider a $D\geq4$ warped-product metric $g_{\mu \nu}$ as in \cref{eq:metricDgeq4}, but I will make more precise assumptions on its properties.
Namely, I will assume that the metric $g_{ab}$ defined on the two-dimensional sections is Lorentzian, and reiterate the assumption that the $(D-2)$-dimensional sections are spheres.
Moreover, I will assume that the spacetime is asymptotically flat, and that both $g_{\mu \nu}$ and $g_{ab}$ have a Killing vector that is timelike in a neighbourhood of spacial infinity --- i.e.~that they are static.
Coordinates can be chosen such that the metric read
\be\label{eq:met_WoH_xgauge}
g_{\mu \nu} \dd{x^\mu} \dd{x^\nu} = - f(x) \dd{v^2} + 2 \mathscr{h}(x) \dd{v}\dd{x} + \phi^2(x) \gamma_{ij} \dd{\theta^i}\dd{\theta^j} \, ,
\ee
I will assume that the metric components are regular, except possibly at true spacetime singularities.

For the sake of this discussion, such a metric will be said to describe a wormhole if the warping function $\phi(x)$ has a (local) minimum --- see e.g.~\cite{MorrisWormholesSpacetime1988,VisserLorentzianWormholes1996,HochbergGeometricStructure1997,LoboExoticSolutions2007,McNuttGeometricSurfaces2021} for further details.
The location $x=x_0$ of said minimum is defined by the conditions
\be\label{eq:throat_def}
\eval{\dv{\phi}{x}}_{x=x_0} = 0
\qq{and}
\eval{\dv[2]{\phi}{x}}_{x= x_0} > 0 \, ,
\ee
and it is usually referred to as the wormhole's \qu{throat} (or \qu{mouth}).
For simplicity, I assume that only one throat be present.
For future convenience, I introduce the parameter $\ell:= \phi(x_0)$, and note that this measures the size of the throat. 

An important remark is in order at this point.
Technically speaking, the requirement that a throat be present is incompatible with the choice of coordinates introduced in \cref{sec:gauge_Birkhoff} and defined by the conditions of \cref{eq:phi_gauge} --- as these require $\nabla_a \phi \neq 0$.
More specifically, such coordinates cannot be chosen in any patch that includes the throat.
However, one is still free to use those coordinates on either side of the throat.
Performing the appropriate coordinate transformation on the metric of \cref{eq:met_WoH_xgauge} leads to identifying
\be\label{eq:h_pole}
h(r) = \mathscr{h}(r)\dv{x}{r} \, .
\ee
This, combined with the conditions of \cref{eq:throat_def}, entails that the metric expressed in the coordinates of \cref{sec:gauge_Birkhoff} is singular at the throat.
Specifically, the limit $x \to x_0^+$ corresponds to $r \to \ell^+$, and in this limit $ h(r) \to + \infty $.
Whether this singularity is physical or a mere coordinate artefact depends on the metric itself.

This remark has important consequences for the reverse engineering of theories whose solutions describe wormholes.
Indeed, according to the discussion of the previous sections, such theories can be written in terms of a function $\eta(\phi)$ that coincides with the function $h(r)$ on shell.
Hence, the Lagrangian of these theories will formally diverge on the surface $\phi = \ell$ of the field space $\{\phi, \chi\}$.
Note that the value of $\ell$ is determined by the Lagrangian in terms of its coupling constants; in particular, $\ell$ is not an integration constant that parametrises different solutions, and therefore it cannot change in result of any physical process --- except, possibly, for renormalisation group running.
Moreover, the solutions of these theories will formally describe the \qu{exterior} of wormhole spacetimes, and it is not entirely obvious that such solutions can be analytically extended past the throat.
This is an important technical caveat, whose resolution I however leave for future work.

Usually, the throat is understood to be a timelike hypersurface, meaning that $f(x_0)>0$. 
In this case, the wormhole is said to be traversable, in the sense that its throat could be crossed in both directions --- at least in principle, i.e.~according to causality criteria alone.
This is the kind of wormholes analysed in the seminal paper by Morris and Thorne \cite{MorrisWormholesSpacetime1988}.

Notably, however, the metric of \cref{eq:met_WoH_xgauge} can describe spacetimes that contain horizons, the outermost of which must be an event horizon as a consequence of asymptotic flatness. 
Such horizons are associated with the zeroes of $f(x)$; hence, in the terminology of this article, their existence is independent from, and perfectly compatible with, that of a wormhole's throat.
Therefore, in particular, it is possible that the throat be a spacelike hypersurface.
This requires $f(x_0)<0$, which entails that such a throat must lie behind an event horizon.
In this case, the throat can only be traversed in one direction, and the wormhole thus resembles a cosmological bounce.
A spacetime of this kind has become known as \qu{black bounce} --- see e.g.~\cite{SimpsonBlackbounceTraversable2019,SimpsonVaidyaSpacetimes2019,MazzaNovelFamily2021,ShaikhConstrainingAlternatives2021,FranzinChargedBlackbounce2021,LoboNovelBlackbounce2021} ---, although the term \qu{black universe} has also been proposed \cite{BronnikovRegularPhantom2006,BronnikovRegularBlack2007,BronnikovBlackUniverses2011}.
Moreover, the limiting case in which the throat is a null hypersurface, i.e.~$f(x_0)=0$, is also admissible.
Such a throat would be traversable in only one direction, similarly to the black bounce case; but it would coincide with a black hole's event horizon. 
A spacetime of this kind has been called \qu{one-way} \cite{SimpsonBlackbounceTraversable2019} or \qu{null} \cite{MazzaNovelFamily2021} wormhole.

Below, I will analyse three explicit examples of wormholes and black bounces, chosen out of infinitely many other possibilities with the aim of showcasing the versatility of the methods presented herein --- as well as their limitations and shortcomings.
For definiteness, I will take $D=4$, although the extension to higher dimensions is straightforward.

    \subsection{The Simpson--Visser black bounce \label{sec:SV}}

A prototypical example of the variety of spacetimes described above is the metric introduced by Simpson and Visser in \cite{SimpsonBlackbounceTraversable2019}:
\be\label{eq:metricSV}
g_{\mu \nu}^\text{SV} \dd{x^\mu} \dd{x^\nu} 
&= - \left( 1 - \frac{2M}{\sqrt{x^2 + \ell^2} } \right) \dd{v^2} + 2 \dd{v} \dd{x} + \left( x^2 + \ell^2 \right) \gamma_{ij} \dd{\theta^i}\dd{\theta^j} \nonumber\\
&= - \left( 1 - \frac{2M}{r} \right) \dd{v^2} + 2 \frac{r}{\sqrt{r^2 - \ell^2}} \dd{v} \dd{r} + r^2 \gamma_{ij} \dd{\theta^i}\dd{\theta^j} \, .
\ee
For $\ell>2M$, this spacetime has no horizon and it describes a wormhole that is traversable in the sense of Morris and Thorne.\footnote{Note that for $M=0$ this metric reduces to that of the so-called Ellis drainhole or Ellis--Bronnikov wormhole \cite{EllisEtherFlow1973,BronnikovScalartensorTheory1973} --- an early example of wormhole metric.}
For $\ell<2M$, instead, the spacetime has an event horizon at $r=2M$ and a spacelike throat hidden behind of it.
For $\ell = 2M$, the throat is null and coincides with the event horizon.
Note that, in any case, the spacetime does not contain any physical singularity, in the sense that it is geodesically complete, and that all the invariants constructed out of the Riemann tensor are bounded functions over their whole domain of definition. 

In light of these considerations, it is most compelling to investigate the reverse engineering of a theory that yields the Simpson--Visser metric as solution.
Applying the strategy outlined in \cref{sec:theorysol}, it is straightforward to see that such a theory must have
\be
\eta(\phi) = \eta_\text{SV}(\phi) &:= \frac{\phi}{\sqrt{\phi^2 - \ell^2}} \, , \label{eq:etaSV} \\
\Omega(\phi, \chi) = \Omega_\text{SV}(\phi, \chi) &:= 2\phi \left( 1 - \chi \frac{\phi^2}{\phi^2 - \ell^2} \right)\, . \label{eq:OmegaSV}
\ee
Correspondingly, according to \cref{eq:Omega_def},
\be
\alpha(\phi, \chi) = \alpha_\text{SV}(\phi, \chi) &:= 
2 \left[ \frac{\sqrt{\phi^2-\ell^2}}{\phi} - \chi \frac{\phi}{\sqrt{\phi^2 - \ell^2} } \left( 1 - 2 \frac{\ell^2}{\phi^2 - \ell^2} \right)\right]
\, , \label{eq:alphaSV}\\
\beta(\phi, \chi) = \beta_\text{SV}(\phi, \chi) &:= - 2 \frac{\phi^2}{\sqrt{\phi^2 - \ell^2}} \, . \label{eq:betaSV}
\ee
Therefore, a two-dimensional Lagrangian whose unique solution is the Simpson--Visser metric is [cf.~\cref{eq:ActHorn_alphabeta}]
\be\label{eq:ActSV}
S_\text{SV} \left[ g_{ab}, \phi \right] &= \int \dd[2]{x} \sqrt{-g} \left\{
2 \left[ \frac{\sqrt{\phi^2-\ell^2}}{\phi} - \chi \frac{\phi}{\sqrt{\phi^2 - \ell^2} } \left( 1 - 2 \frac{\ell^2}{\phi^2 - \ell^2} \right)\right]
\right. \nonumber\\
&\phantom{=} \left.
- 4 \frac{\phi^2}{\sqrt{\phi^2 - \ell^2}} \Box{\phi}
+ \left[ \phi \sqrt{\phi^2 - \ell^2} + \ell^2 \arctan(\frac{\phi}{\sqrt{\phi^2 - \ell^2}}) \right]\mathcal{R}
\right\} \, .
\ee
This action can then be uplifted to four dimensions following the strategy outlined in \cite{BorissovaAll$2D$2026}.
However, describing such an uplift would require introducing a considerable amount of rather heavy notations, arguably without adding substantial understanding.
It thus appears more interesting to continue the discussion with a series of salient remarks concerning the results so far obtained.

First of all, observe that $\ell$ appears as a parameter at the level of the action, and it therefore plays the role of a coupling constant.
All the expressions of \cref{eq:etaSV,eq:OmegaSV,eq:alphaSV,eq:betaSV,eq:ActSV} reduce to their counterparts arising from the symmetric reduction of four-dimensional general relativity, upon taking $\ell\to0$ --- cf.~\cref{eq:OmegaGR,eq:alphabetaGR,eq:ActGR}.
This is expected, but represents an important sanity check nonetheless.
Note however that the Lagrangian is formally divergent at $\phi = \ell$, as anticipated.

Further note that $\alpha_\text{SV}$ and $\beta_\text{SV}$ cannot be written as in \cref{eq:alpha_NoDer,eq:beta_NoDer}, meaning that the uplift of the action in \cref{eq:ActSV} to higher dimensions must involve derivatives of the Riemann tensor.
This fact notwithstanding, such an uplift can be considered a deformation of general relativity.
Recalling the discussion at the end of \cref{sec:int2D}, the reader might recognise one of the two deformations introduced there --- specifically, the second.
Namely, the Simpson--Visser metric can be obtained, formally, as a deformation of the Schwarzschild metric whereby $f(r)$ is left unchanged and $h(r)$ is modified appropriately. 
Indeed, one may verify that $\alpha_\text{SV}$ and $\beta_\text{SV}$ can be obtained from $\alpha_\text{GR}$ and $\beta_\text{GR}$ employing the relations of \cref{eq:alpha_deform,eq:beta_deform}.

    \subsection{A singular deformation \label{sec:sing_deform}}

The last remark in the previous subsection motivates the investigation of the other (i.e.~the first) deformation proposed at the end of \cref{sec:int2D}.
Namely, I will consider a theory that is generated by the same potential $\Omega_\text{GR}$ as general relativity, but differs from this due to a different choice of $\eta(\phi)$.

As explained in \cref{sec:int2D}, the solution of such a theory has the same profile of $\chi(r)$ as general relativity.
The resulting metric, however, differs from the Schwarzschild metric, since it reads
\be\label{eq:metricSing}
g_{\mu \nu}^\text{sing.} \dd{x^\mu} \dd{x^\nu} = - \left( 1 - \frac{2M}{r} \right) h^2(r) \dd{v^2} + 2 h(r) \dd{v}\dd{r} + r^2 \gamma_{ij} \dd{\theta^i}\dd{\theta^j}\, ,
\ee
where as usual $h(r) = \eta(r)$.

It is quite interesting to note that, contrary to the previous example, this metric generically displays a true spacetime singularity at the throat, quite irrespective of the choice of $\eta$.
Indeed, the two-dimensional Ricci scalar for the metric of \cref{eq:metricSing} reads
\be
\mathcal{R} = - 2 \left( 1- \frac{2M}{r} \right) \frac{1}{h(r)} \dv[2]{h(r)}{r} + \frac{2M}{r^3} \left( 2 - \frac{3r}{h(r)} \dv{h(r)}{r} \right) \, ,
\ee
while its four-dimensional analogue is
\be
R^{(4)} = \mathcal{R} - \frac{4}{rh(r)} \dv{}{r} \left[ \left(1-\frac{2M}{r}\right) h(r) \right] + \frac{4M}{r^3} 
\ee
--- cf.~\cref{eq:RicciDgeq4} and \cite{BorissovaAll$2D$2026}.

At the throat, $h(r)$ must diverge, according to the discussion below \cref{eq:h_pole}.
That the derivatives of $h(r)$ should also diverge is quite intuitive, and it can be shown explicitly.
Indeed, referring to the notation of \cref{eq:met_WoH_xgauge}, one can show
\be
\dv{h(r)}{r} &= - h^3(r) \dv[2]{\phi}{x} \, ,\\
\dv[2]{h(r)}{r} &= h^4(r) \left[ 3 h(r) \left( \dv[2]{\phi}{x} \right)^2 - \dv[3]{\phi}{x} \right] \, ,
\ee
and so on.
Hence, according to \cref{eq:throat_def}, 
\be
\lim_{r \to \ell^+} \dv{h(r)}{r} = - \infty \, ,
\qq{with} 
\lim_{r \to \ell^+} \frac{1}{h^3(r)} \dv{h(r)}{r} < \infty 
\qq{and negative;} 
\ee
and similarly for $\dv*[2]{h(r)}{r}$ and higher derivatives.
Therefore, barring fortuitous cancellations, both the two- and four-dimensional Ricci scalars diverge as $r\to\ell^+$.
(They also diverge as $r\to0$, but this point is supposed to lie outside of the domain of validity of this coordinate.)

The existence of this singularity is not surprising in itself, and in fact I have already hinted to this possibility before.
Still, this example offers the opportunity to stress that the approach described here applies to a wide set of spacetimes, quite irrespective of their regularity.

One might wonder what conditions a higher-dimensional theory ought to satisfy in order to yield regular solutions.
This question would definitely deserve further scrutiny, which I leave for future work.
Here, I only point out that these conditions would presumably constitute a rephrasing of the regularity conditions for four- and higher-dimensional metrics --- which are fairly well understood, see e.g.~\cite{MazzaHeartDarkness2023,Carballo-RubioNonsingularParadigm2025} and references therein.
Further note that this question has been partially addressed in \cite{BoyanovRegularVaidya2025} for black holes.

    \subsection{A wormhole without a theory \label{sec:DS}}

As a third and last example of static wormhole spacetime, I shall consider the metric proposed by Damour and Solodukhin \cite{DamourWormholesBlack2007}.
In coordinates coherent with \cref{sec:gauge_Birkhoff}, this metric reads
\be\label{eq:metricDS}
g_{\mu\nu}^\text{DS} \dd{x^\mu} \dd{x^\nu} = - \left( 1 + \lambda^2 - \frac{2M}{r} \right) \dd{v^2} + 2 \sqrt{ \frac{r(1+\lambda^2) - 2M}{r - 2M} } \dd{v}\dd{r} + r^2 \gamma_{ij} \dd{\theta^i} \dd{\theta^j}\, .
\ee
The throat is located at $r=2M$, hence according to the notations introduced above one should identify $\ell = 2M$.
The (Komar) mass, on the other hand, should be identified with the combination $M/\sqrt{1+\lambda^2}$.
The notation of \cref{eq:metricDS} is therefore somewhat unfortunate, but it is standard.

At any rate, what makes this example worth analysing is the fact that both the mass and the throat's size, or equivalently the parameters $M$ and $\lambda$, enter \emph{both} components of the two-dimensional metric.
This poses a problem when reverse engineering a theory that yields \cref{eq:metricDS} as solution.

Indeed, the procedure outlined in \cref{sec:theorysol} leads to a clear distinction between a parameter that emerges as an integration constant, and one or more parameters which appear in the solution because they enter the action as coupling constants. 
It is most natural to relate the integration constant to the mass of the spacetimes, since this is an extensive quantity that is assumed to characterise any spacetime.
This is exactly what happens for general relativity as well as, for instance, for the regular black holes of \cite{BorissovaRegularBlack2026a}, and for both of the examples above.
Crucially, the reverse engineering of \cref{sec:theorysol} applies to metrics as in \cref{eq:metric2D} [or \cref{eq:metricDgeq4}] such that $h(r)$ is independent from said integration constant.
If either $M$ or $\lambda$ are to be interpreted as an integration constant, this is not the case for the metric of \cref{eq:metricDS}.

If one were to insist in attempting a reverse engineering as described before, one might be lead to the theory determined by
\be
\eta(\phi) = \eta_\text{DS} (\phi) &= \sqrt{\frac{\phi(1+\lambda^2) - 2M}{\phi-2M}} \, ,\\
\Omega(\phi, \chi) = \Omega_\text{DS}(\phi, \chi) &= 2\phi \left( 1 +\lambda^2 - \chi\eta^2 \right) \, .
\ee
The equations of motion entail that, on shell, $\Omega_\text{DS}$ evaluates to a constant, which I shall call $4 \Tilde{M}$.
The resulting solution reads
\be
g_{\mu\nu} \dd{x^\mu} \dd{x^\nu} = - \left( 1 + \lambda^2 - \frac{2\tilde{M}}{r} \right) \dd{v^2} + 2 \sqrt{ \frac{r(1+\lambda^2) - 2M}{r - 2M} } \dd{v}\dd{r} + r^2 \gamma_{ij} \dd{\theta^i} \dd{\theta^j}\, ,
\ee
and it coincides with that of \cref{eq:metricDS} only for the specific value $\tilde{M} = M$.
For other values of $\tilde{M}$, the metric still describes a wormhole or a black bounce, but strictly speaking not of the Damour--Solodukhin kind.

This kind of fine tuning is quite ubiquitous in other approaches to the reverse-engineering problem, for instance in the context of non-linear electrodynamics --- see e.g.~\cite{Ayon-BeatoBardeenModel2000,BronnikovRegularMagnetic2001,BalartRegularBlack2014,FanConstructionRegular2016,BronnikovNonlinearElectrodynamics2018,BronnikovRegularBlack2023}.
The procedure discussed in this article allows to circumvent this problem in several, but evidently not all cases.
However, I wish to stress that these limitations do not entail that the Damour--Solodukhin wormhole cannot arise as a solution of a higher-dimensional purely metric theory.
On the contrary, virtually any metric can arise in this way.
What this example shows is that not all metrics can arise from (extended) quasi-topological gravities via the reverse engineering described in \cref{sec:theorysol}.

    \section{Dynamic wormholes and black bounces \label{sec:dynamicWoH}}    
    
The previous section has dealt with the reverse engineering of static wormholes and black bounces.
Here, I wish to direct the discussion on to dynamical spacetimes.
As mentioned in \cref{sec:gauge_Birkhoff}, all two-dimensional Horndeski theories satisfy a Birkhoff--Jebsen theorem and their solutions are therefore automatically static.
Hence, in order to allow for a time dependence, one has to move beyond that framework, for instance by introducing minimally coupled matter sources.

Specifically, I will consider higher-dimensional theories constructed out of the metric and a collection of matter fields, and assume that their symmetric reduction gives rise to two-dimensional actions of the form
\be\label{eq:ActMatter}
S \left[ g_{ab}, \phi, \Psi \right] = S_\text{2D Horndeski} \left[ g_{ab}, \phi \right] + S_\text{matter} \left[ g_{ab}, \phi, \Psi \right] \, ,
\ee
where $\Psi$ collectively denotes the matter fields.
This setting has been discussed in \cite{BoyanovRegularVaidya2025,Carballo-RubioMasterField2026,ArrecheaEffectiveGeometrostatics2026}, to which I refer the reader for further details.
The equations of motion derived from varying the action of \cref{eq:ActMatter} with respect to the metric and the scalar field can be written, schematically, as
\be
\mathcal{E}_{ab} &= T_{ab} \, ,\\
\mathcal{E} &= \mathcal{T} \, ,
\ee
where $\mathcal{E}_{ab}$ and $\mathcal{E}$ are as in \cref{eq:Eab,eq:Ephi} while $T_{ab}$ and $\mathcal{T}$ are defined appropriately.
These should be complemented with the equations of motion for the matter fields $\Psi$, which I assume to be satisfied.
As for the vacuum case, the equations of motion are related by a Bianchi identity \cite{JacobsonInitialValue2011} which, assuming the matter fields are on shell, reads
\be
\nabla^a \left( \mathcal{E}_{ab} - T_{ab}\right) + \frac{1}{2} \left( \mathcal{E} - \mathcal{T} \right) \nabla_b \phi = 0 \, .
\ee
This is the analogue of \cref{eq:BianchiID}; as before, I assume that generically $\nabla_a \phi \neq 0$, so that the tensorial equations imply the scalar equation.

Here, I follow the strategy presented in \cite{BoyanovRegularVaidya2025,FrolovVaidyaTypeSolutions2026} and consider Vaidya-like solutions \cite{VaidyaExternalField1943,VaidyaExternalField1999,VaidyaGravitationalField1951,VaidyaGravitationalField1999,VaidyaNonstaticSolutions1951,WangGeneralizedVaidya1999}.
That is, I consider a source of the form
\be\label{eq:TabVaidya}
T_{ab} = 2 \dv{M(v)}{v} \partial_a v \, \partial_b v \, ,
\ee
with $M(v)$ and arbitrary function of $v$.
Note that this tensor is proportional to the symmetric reduction of the familiar source of the $D$-dimensional Vaidya spacetime \cite{IyerVaidyaSolution1989}, i.e.~to the stress--energy tensor of \qu{null dust}.
Further note that, by construction, the only static solutions are the \qu{vacuum} solutions of the Horndeski theory, i.e.~this source does not give rise to \qu{hairy} static solutions. 

In the gauge of \cref{sec:gauge_Birkhoff}, the equations of motion become
\be
\alpha_\chi - \beta_\phi - \beta \frac{\partial_r h}{h} &= 0  \, , \label{eq:ETrr} \\
- \frac{1}{2} \left( \alpha h + \beta h \partial_r \chi \right) &= 0 \, , \label{eq:ETvr} \\
\frac{1}{2} \beta h \partial_v \chi &= 2 \dv{M(v)}{v} \, . \label{eq:ETvv}
\ee
Note that the first two equations are identical, respectively, to \cref{eq:Err,eq:Evr}, while the last equation rules the time evolution of the kinetic term.
These equations can be solved in a way that is completely analogous to the one described in \cref{sec:theorysol}.
Supposing the two-dimensional Horndeski theory belongs to the (extended) integrable class, i.e.~that $\alpha$ and $\beta$ satisfy the (extended) integrability condition of \cref{eq:ext_int_cond}, there will exist $\eta(\phi)$ and $\Omega(\phi, \chi)$ such that the writing of \cref{eq:Omega_def} holds.
In such a case, \cref{eq:ETrr,eq:ETvr} are equivalent to
\be
\frac{\partial_r h(r)}{h(r)} &= \frac{\eta_\phi (r)}{\eta(r)} \, , \\
\pdv{\Omega}{r} &= 0 \, ,
\ee
exactly as in the static case; while \cref{eq:ETvv} is equivalent to
\be
\pdv{\Omega}{v} = \dv{}{v} 4M(v) \, .
\ee
This equation, in turn, can be integrated straightforwardly to give
\be
\Omega = 4 M(v) + M_0 \, ,
\ee
with $M_0$ an integration constant, which from now on I will absorb in a redefinition of $M(v)$.
The solution can therefore be written as in \cref{eq:metric2D} with $h$ and $f$ determined implicitly, up to redefinitions of time, by
\be
h(v,r) = \eta(r)
\qq{and}
\Omega\left( r,\chi(v,r) \right) = 4M(v) \, .
\ee

As an example, one could apply this formalism to the Simposon--Visser metric introduced in \cref{sec:SV}.
This entails considering the two-dimensional Horndeski theory that yields the Simpson--Visser metric as a static solution, and coupling it to the matter source of \cref{eq:TabVaidya}.
The resulting solution, uplifted to higher dimensions, reads
\be
g_{\mu\nu}^\text{SV--Vaidya} \dd{x^\mu} \dd{x^\nu} = 
- \left( 1 - \frac{2M(v)}{r} \right) \dd{v^2} + 2 \frac{r}{\sqrt{r^2- \ell^2}} \dd{v}\dd{r} + r^2 \gamma_{ij} \dd{\theta^i} \dd{\theta^j}\, , 
\ee
which is identical to the metric of \cref{eq:metricSV} except for the replacement $M \mapsto M(v)$.
This Vaidya-like extension of the Simpson--Visser metric has been introduced and analysed in \cite{SimpsonVaidyaSpacetimes2019}: 
it describes a wormhole with a throat located at $r=\ell$, and an apparent horizon located at $r=2M(v)$.
Hence, depending on the shape of $M(v)$, this spacetime should be interpreted as a traversable wormhole or a black bounce that is accreting or radiating mass; notably, the metric can describe transitions between a traversable wormhole and a black bounce, and vice versa.
However, it does not describe a dynamical wormhole throat of the kind described in e.g.~\cite{HaywardDynamicWormholes1999,HochbergDynamicWormholes1998,HochbergNullEnergy1998,TomikawaNewDefinition2015}.
In this article, this metric is derived as a solution to a well-defined higher-dimensional metric theory of gravity coupled to matter.

It is worth noting that, in this example, the time evolution only affects the value of the parameter $M$ --- and not of $\ell$.
This feature is a direct consequence of the construction herein described, and it applies to virtually any application of it.
Indeed, if the vacuum solutions are the \emph{only} static solutions, they can be considered as \qu{equilibrium states}, which dynamical transients may connect;
since different vacuum solutions are labelled by the value of the integration constant $M$, in practice, the dynamics can only affect its value.

This entails that, as I have noted several times, in this framework the dynamics can never affect the value of the parameter $\ell$, which should be interpreted as a coupling constant.
In particular, $\ell$ cannot be generated dynamically, e.g.~during gravitational collapse.
Hence, in this context, wormholes cannot be \qu{formed}, at least not in the same sense in which, say, black holes are.

    \section{Conclusions\label{sec:concl}}

In this article, I have extended previous results on quasi-topological gravities in four and higher dimensions, and on the corresponding two-dimensional \qu{integrable} Horndeski theories obtained by dimensionally reducing the former on warped product spacetimes.
Such an extension encompasses situations in which the metric on the two-dimensional portion of the warped product is described in terms of two functions of one variable --- instead of the single function considered in most analyses.
Hence, in particular, it allows to account for wormhole spacetimes.

The result of this extension is a simple and straightforward procedure to reverse engineer, starting from (almost) any static two-dimensional metric, a two-dimensional Horndeski theory having said metric as a solution; and to uplift to $D\geq4$ dimensions, respectively, the two-dimensional metric to a warped-product metric with symmetric sections, and the Horndeski theory to a modified gravity theory constructed out of the $D\geq4$ metric, the associated Riemann tensor, and derivatives thereof.
Such a reverse engineering thus provides an \qu{effective} geometrodynamics for many spacetimes of interest for phenomenology --- such as non-singular black holes, wormholes and black bounces, and other black hole \qu{mimickers}.

Notably, the theories obtained through this procedure automatically enjoy a Birkhoff--Jebsen theorem.
Moreover, their solutions depend continuously on a (single) parameter, which arises as an integration constant when solving the equations of motion, and can therefore be considered a primary hair; this parameter is thus most naturally associated with the (Komar) mass of the spacetime.
Hence, in this framework, any other parameter that might appear in the solution must be a combination of parameters already present at the level of the action, i.e.~it is tantamount to a coupling constant: this is the case, for instance, for the regularisation scale of non-singular black holes, and for the size $\ell$ of wormholes' throats.
In essence, this statement constitutes a no-hair theorem, and finding exceptions thus requires circumventing said theorem --- for example by introducing matter sources, see \cite{Carballo-RubioChargingRegular2026}. 

This feature determines a technical limitation in the applicability of this procedure.
Indeed, in order for the reverse engineering to work properly, the parameter that is to be interpreted as the integration constant must appear in only one of the two metric function --- $f(r)$, in the notation of the text --- and not in the other --- viz.~$h(r)$. 
If this is not the case, the procedure becomes somewhat ambiguous, insofar as it might require the fine tuning of the integration constant with respect to the other parameters.
Such a fine tuning is rather ubiquitous in other approaches to the reverse-engineering problem, such as those based on non-linear electrodynamics.
It is quite remarkable, therefore, that the procedure described herein allows to avoid the issue in many --- possibly most, but definitely not all --- relevant cases.

To showcase the versatility, as well as the limitations, of this procedure, I applied it to three prototypical examples of static and spherically symmetric wormholes --- see \cref{sec:staicWoH}.
The first of such examples is the Simpson--Visser black bounce \cite{SimpsonBlackbounceTraversable2019}: this is a case to which the reverse engineering applies perfectly, yielding a theory which, though unwieldy, can be considered a deformation of general relativity. 
The second example is another metric to which the reverse engineering applies without issues, but which differs qualitatively from the Simpson--Visser black bounce inasmuch as the spacetime is singular.
Finally, the third example is the Damour--Solodukhin wormhole \cite{DamourWormholesBlack2007}, a metric that depends on two parameters which enter both components of the two-dimensional metric: the reverse engineering therefore does not apply as neatly to this case, and the discussion gives an account of the ensuing difficulties.

An interesting remark concerning the application of this procedure to spacetimes describing wormholes is that the reverse-engineered Lagrangian displays a singularity in field space, associated, roughly speaking, to the throat's size $\ell$.
Notably, this singularity is not related to the regularity, or lack thereof, of the metric itself --- as testified by the explicit examples of \cref{sec:staicWoH}.
Rather, it is a manifestation of the so-called \qu{flare-out} conditions that characterise a wormhole's throat.
These conditions are realised, in the coordinates used in this article, as a divergence in one of the metric components --- namely, $h(r)$.

This remark points to another technical limitation of the approach adopted herein, specifically for what concerns wormholes.
The coordinates introduced in \cref{sec:gauge_Birkhoff} and used throughout do not cover the throat itself; rather, they are viable coordinates only on either side of the throat.
Hence, strictly speaking, the wormhole metrics considered, for example, in \cref{sec:staicWoH} are solutions of the corresponding reverse-engineered theory only away from the throat, on either side of it.
Naively, this does not appear to be a problem since, unless the throat is truly singular, there exist coordinates in which the metric is analytic there; and the analytic continuation of a solution to the equations of motion should remain a solution, according to general theorems on partial differential equations.
The fact that the reverse-engineered Lagrangian is formally not analytic, however, renders this deduction not quite straightforward, and definitely worth of further investigation.

The issue might be best described with an example.
Consider a wormhole obtained by performing a \qu{surgery} on the Schwarzschild metric, i.e.~by gluing two mirror copies of the Schwarzschild spacetime at some $r=\ell$. 
In general relativity, such a gluing requires the introduction of a thin shell of (exotic) matter localised at the throat --- see e.g.~\cite[ch.~15]{VisserLorentzianWormholes1996}.
Away from the throat, employing the reverse engineering described here yields general relativity, since the solution there is Schwarzschild; by naive analyticity arguments, one would be led to extend the Schwarzschild solution, thereby never accounting for the throat --- i.e.~for the non-trivial global topology of the manifold.
These arguments might be excessively naive, however, as the treatment of thin shells and junction conditions in these theories is not entirely trivial --- cf.~\cite{FilippoMassInflation2026}.

At any rate, although these limitations definitely call for a supplement of inquiry, the formalism presented in this article is extremely powerful and versatile, and it readily lends itself to further developments. 
A simple example of such developments has been presented in \cref{sec:dynamicWoH}, where I investigated the possibility of circumventing the Birkhoff--Jebsen theorem by coupling the metric to a Vaidya-like matter source --- along the lines of \cite{BoyanovRegularVaidya2025}.
This coupling gives rise to time-dependent solutions describing accreting or radiating compact objects, such as wormholes.
In particular, I worked out explicitly the example of the Simpson--Visser black bounce, deriving its Vaidya-like extension as a solution to a well-defined higher-dimensional metric theory of gravity coupled to a matter source whose stress--energy tensor is that of a null dust.
Interestingly, however, these spacetimes are not proper dynamical wormholes, in the sense that their throat is not dynamical.
This result is yet another manifestation of the fact that, in this framework, the throat's size $\ell$ is fixed by coupling constants and is, in essence, a fundamental scale.
Hence, in particular, in this setting wormholes appear impossible to \qu{form}, in the sense that a throat cannot \qu{open up} as a result of, say, gravitational collapse.

Clearly, however, these explorations are preliminary and simplistic, and their result should be taken as an invitation for further investigations.

    \section*{Acknowledgments}

I would like to thank Raúl Carballo-Rubio for his encouragement and the countless stimulating discussions.
The author acknowledges support of ANR grant StronG (ANR-22-CE31-0015-01). 


\bibliographystyle{JHEP}
\bibliography{biblio.bib}

\end{document}